\documentclass[
 reprint,
 amsmath,amssymb,
 aps,twocolumn,
prb
]{revtex4-1}

\usepackage{hyperref}
\usepackage{dcolumn}
\usepackage{amsbsy}
\usepackage{amsmath}

\usepackage{amssymb}
\usepackage{braket}
\usepackage{color}
\usepackage{fullpage}
\usepackage{graphicx}
\usepackage{bm}

\usepackage{subfigure}
\usepackage{hyperref}
\usepackage{cleveref}

\usepackage{epstopdf}

\usepackage{epsfig}

\usepackage{multirow}
\usepackage{diagbox}
\usepackage{graphicx}
\usepackage{dcolumn}
\usepackage{bm}
\usepackage{hyperref}
\usepackage[mathlines]{lineno}
\usepackage{xcolor}
\usepackage[english]{babel}

\def\<{\langle}
\def\>{\rangle}

\def\nn{\nonumber}

\def\beq{\begin{equation}}
\def\eeq{\end{equation}}

\newcommand{\bea}{\begin{eqnarray}}
\newcommand{\eea}{\end{eqnarray}}

\newcommand{\pt}{$\mathsf{PT~}$}

\def\ket#1{| \,#1\, \rangle}
\def\bra#1{\langle \,#1\, |}
\def\scx#1#2{\langle \,#1\, |\, #2\, \rangle}

\def\me#1#2#3{\langle \,#2\, |\,#1\,|\, #3\,\rangle}

\def\dfrac#1#2{{\displaystyle\frac{#1}{#2}}}

\def\lsim{\mathrel{\rlap{\lower4pt\hbox{\hskip1pt$\sim$}}
    \raise1pt\hbox{$<$}}}         
\def\gsim{\mathrel{\rlap{\lower4pt\hbox{\hskip1pt$\sim$}}
    \raise1pt\hbox{$>$}}}         

\begin{document}

\title{
Simulating non-Hermitian dynamics of a multi-spin quantum system  \\ and an emergent central spin model}

\author{Anant V. Varma}%
 \email{anantvijay.cct@gmail.com}
 
\author{Sourin Das}
 \email{sourin@iiserkol.ac.in , sdas.du@gmail.com}

\affiliation{%
Indian Institute of Science Education \& Research Kolkata,
Mohanpur, Nadia - 741 246, 
West Bengal, India
}%
\begin{abstract}
{\color{black} 
It is possible to simulate the dynamics of a single   spin-$1/2$ (\pt   symmetric) system  by conveniently
 embedding it into a subspace of a larger Hilbert space  with unitary dynamics.  Our goal is to formulate a many body generalization of this idea i.e., embedding many body non-Hermitian dynamics. As a first step in this direction, we investigate  embedding  of ``$N$" non-interacting  spin-$1/2$ (\pt symmetric) degrees of freedom, thereby unfolding the complex nature of such an embedding procedure. It turns out that the resulting Hermitian Hamiltonian represents a cluster of $N+1$ spin halves with ``all to all", $q$-body interaction terms ($q=1, \ldots,N+1$) in which the  additional spin-$1/2$ is a part of the  larger embedding space. We can visualize it as a strongly correlated central spin model with the additional spin-$1/2$  playing the role of central spin. We find that due to the orthogonality catastrophe, even a vanishing small exchange field applied along the anisotropy axis of the central spin leads to a strong suppression of its decoherence arising from spin-flipping perturbations.


}
\end{abstract}

                              \maketitle
\section{\label{intro}Introduction}
{\color{black}

\pt symmetric non-Hermitian systems have been a topic of  interest ever since it was discussed by Bender et. al~\cite{CMB,PDorey,CBend,DCBro}.
Some of the  striking features of such systems are:  extreme acceleration of state evolution~\cite{Jones2007}, quantum state discrimination~\cite{Royal2013}, perfect quantum state transfer~\cite{Zsong},  unidirectional optical transmission  and single mode lasing~\cite{Rame,HCao,JWong,NMois,WChen,UHass}, anomalous and unconventional states in many-body systems~\cite{Shio,Yuto,Hama} etc. Naturally, it would be desirable  to harness some of these exotic features  for practical use.  For instance, exploiting accelerated state evolution of \pt symmetric quantum systems for faster quantum information processing would serve as a valuable resource. However, physical realization of \pt symmetric quantum systems is challenging.

There has been remarkable progress in realizing physical systems which demonstrate  \pt symmetry~\cite{NChris,LFeng,GMak,AGuo,GMark,Tang} including the many-body quantum systems~\cite{Li,Klauck}. A concrete idea of realizing \pt symmetric quantum dynamics was discussed by G\"{u}nther and Samsonov~\cite{Gunther}. Simulation of  the non-Hermitian dynamics governed by a \pt symmetric Hamiltonian was achieved by embedding the dynamics of these states in a subspace of a higher dimensional Hilbert space such that the total system evolves via unitary dynamics governed by Hermitian Hamiltonian. Finally, the \pt symmetric state at each instant of time can be obtained by performing  appropriate projective measurements on the subspace. In particular, the authors embedded non-Hermitian dynamics of \pt symmetric spin-$1/2$ residing in $\mathbb{C}^{2}$ by using a larger Hermitian Hamiltonian defined on $\mathbb{C}^{4}$. And, it is encouraging to note that the experimental implementation of these ideas has been reported very recently in Ref.~\cite{Peng}.

 Embedding of higher dimensional non-Hermitian systems has also been discussed by Kawabata et. al~\cite{Ueda} and others~\cite{Ray,Kumar} but,  the possibility of embedding a many-body \pt  symmetric system, which is interesting from the perspective of quantum many-body physics, has not been carried out earlier. In the present work, we explore this possibility and show that embedding a collection of free \pt symmetric spin halves leads to complex embedding Hamiltonian which corresponds to a cluster of strongly correlated spin halves with anisotropic non-local interactions. As an  offshoot, we obtain an exactly solvable central spin model with  complex anisotropic interaction.
The primary goal of  this article is to study and understand the complexity of embedding many-body non-Hermitian dynamics using ideas explored in Ref.~\cite{Gunther,Ueda,Ray,Kumar}.

 The paper is structured as follows. We describe our formalism of embedding non-Hermitian dynamics in Sec.~\ref{embedding}.  We then apply this formalism to the case of ``$N$ free \pt symmetric spins" and obtain the embedding Hamiltonian which is shown to be comprising of of ``all to all", $q$-body interaction terms where $q=1,...N+1$ in Sec.~\ref{embedding N}.  We explicitly study the $N=2$ case  in Sec.~\ref{embedding N=2},  which corresponds to one ancilla (central) spin coupled to two other bath spins described by a Hamiltonian  which includes $3$-body interaction term that can get as large as the single body terms hence indicating presence of strong correlation.   In Sec.~\ref{entanglement}, we study the eigenfunctions of embedding Hamiltonian and discuss the entanglement encoded in these states.  Finally, in Sec.~\ref{central-spin}, we study the embedding Hamiltonian from the perspective of the central spin model and discuss the dark and bright states of the central spin. We conclude in Sec.~\ref{discussion}. 

\section{\label{embedding}  Embedding non-Hermitian dynamics}
%
The necessary and sufficient condition for an operator $\hat{H}_{PT}$  to have real eigenvalues is the existence of a positive-definite matrix $\hat{\eta}$ such that~\cite{Jordan1969},
\begin{equation}
\hat{\eta} \  \hat{H}_{PT}^{} = \hat{H}_{PT}^{\dagger} \ \hat{\eta} ~.  
\label{E00}
\end{equation}
Furthermore, both \pt symmetric non-Hermitian Hamiltonian and Hermitian Hamiltonian share real-valued eigenspectrum which implies the existence of a similarity transformation mapping of the \pt  symmetric Hamiltonian onto to a Hermitian one   (henceforth, referred to as  the seed Hamiltonian, {{$\hat{h}$}}) given by
\begin{equation}
   \hat{H}_{PT} ^{} = \hat{\mathcal{S}} \  \hat{h}  \ \hat{\mathcal{S}}^{-1}~,
\label{E0}
\end{equation}
where $\hat{\mathcal{S}}$ is a complex invertible matrix and is unique up to a unitary transformation acting on $\hat{h}$. We will see later that this identification of $\hat{h}$ for a given $ \hat{H}_{PT}^{}$ plays a crucial role in formulation of the problem in the following sense - this helps us in obtaining a natural interpretation of the outcome of embedding in terms of the local Hermitian degrees of freedom which constitute the many-body Hermitian Hamiltonian $\hat{h}$ which is why it is called the seed Hamiltonian. 

Eq.~\ref{E00} and Eq.~\ref{E0} together implies that  $\hat{\eta} = (\hat{\mathcal{S}}\hat{\mathcal{S}}^{\dagger})^{-1}$. Now we use the polar decomposition: $ \hat{\mathcal{S}} = \hat{\mathcal{P}} \  \hat{\mathcal{U}}$ where the matrix $\hat{\mathcal{P}}$ is a positive semi-definite Hermitian matrix and $\hat{\mathcal{U}}$ is a unitary matrix. This decomposition is unique for invertible matrix $\hat{\mathcal{S}}$ and hence we can identify a relation between  
$\hat{\eta}$ and $\hat{\mathcal{P}}$ given by $\hat{\eta} = ( \hat{\mathcal{S}}^{\dagger})^{-1}  \hat{\mathcal{S}}^{-1} =  \hat{\mathcal{P}}^{-2}$. Henceforth, we drop the unitary operator $\hat{\mathcal{U}}$ without any loss of generality, since it is a global unitary operator. 
%
%
Now we introduce the embedding of a non-Hermitian  \pt  symmetric ${\cal N}\times {\cal N}$  matrix into a larger Hilbert space which is constructed by adding a two level system to the existing one which we will call the ancilla degree of freedom from now on-wards. Following the method of embedding mentioned in~\cite{Ueda}, we start out by writing the total embedding Hamiltonian defined over the extended Hilbert space $\mathcal{H} = \mathcal{H}_{2}\ \otimes \ \mathcal{H}_{\cal N}$  given by
\begin{equation}
    \hat{H}_{T} = \mathbb{I}_{2\times 2} \otimes \hat{\mathcal{A}} + \sigma_{y} \otimes \hat{\mathcal{B}}~,
\label{E1}
\end{equation}
where the operators $\hat{\mathcal{A}}$ and $\hat{\mathcal{B}}$ act on the space of \pt symmetric degrees of freedom while the $ \mathbb{I}_{2\times 2}$ and $ \sigma_{y}$ acts on the space of the extra two-level system introduced for the purpose of embedding. The explicit form the operators $\hat{\mathcal{A}}$ and $\hat{\mathcal{B}}$ in terms of the  $\hat{H}_{PT}^{}$  can be identified as:
\bea
    \hat{\mathcal{A}}&=& \dfrac{1}{c}  \left(\hat{H}_{PT}^{} \hat{\mathcal{Q}}^{-1}+  \hat{\mathcal{Q}} \hat{H}_{PT}^{}\right) \hat{\eta}^{-1} \hat{\mathcal{Q}}~,
\nn\\
    \hat{\mathcal{B}}&=& \dfrac{i}{c}  \left(\hat{H}_{PT}^{} -  \hat{\mathcal{Q}} \hat{H}_{PT}^{} \hat{\mathcal{Q}}^{-1} \right) \hat{\eta}^{-1} \hat{\mathcal{Q}}~. 
\label{E1p}
\eea
These can be further simplified and expressed in terms  of the seed Hamiltonian   %
 \bea
\hat{\mathcal{A}} &=& \dfrac{1}{c} \hat{\mathcal{P}} \left(  \hat{h}  + \hat{\mathcal{Q}}  \hat{h} \ \hat{\mathcal{Q}} \right) \hat{\mathcal{P}}~,
\nn\\
 \hat{\mathcal{B}}&=&\dfrac{i}{c}  \hat{\mathcal{P}} \left[\hat{h}, \hat{\mathcal{Q}} \right] \hat{\mathcal{P}} ~,
\label{E2}
\eea
where   $\hat{\mathcal{Q}}$  is given by %
 \begin{equation}
\hat{\mathcal{Q}}=\left( c \  \hat{\mathcal{P}}^{-2} - \mathbb{I}_{{\cal N} \times {\cal N}} \right)^{1/2} .
\label{E2a}
\end{equation}
Here the constant $c$ is set to be the sum of inverse eigenvalues of the operator $\hat{\mathcal{P}}$ so that the operator $\hat{\mathcal{Q}}$ always stays Hermitian which is essential for keeping $\hat{H}_{T}$ Hermitian. 
%
To understand the above construction of the embedding Hamiltonian, we need to have a careful look at the time evolution of a specific type of wave function which facilitates the simulation of the non-Hermitian \pt symmetric dynamics. We consider the following two wave functions in $\mathcal{H}$ 
\bea
\ket{\Psi_{T}}_{+} &=& \dfrac{1}{\sqrt{c}} \left(  \ket{\uparrow}_{z} \otimes \ket{\psi}_{PT}^{} + \ket{\downarrow}_{z} \otimes  \hat{\mathcal{Q}} \ket{\psi}_{PT}^{} \right)~, \nn\\
\ket{\Psi_{T}}_{-} &=& \dfrac{1}{\sqrt{c}} \left(  \ket{\downarrow}_{z} \otimes \ket{\psi}_{PT}^{} - \ket{\uparrow}_{z} \otimes  \hat{\mathcal{Q}} \ket{\psi}_{PT}^{} \right)~, \nn
\label{E2aa}
\eea
%
where the state $\ket{\psi}_{PT}$ lives in $\mathcal{H}_{N}$  and has a well-defined time evolution under the influence of $\hat{H}_{PT}$  and is normalized with respect to the  \pt symmetric inner product, i.e, $_{PT}\me{\hat \eta}{\psi}{\psi}  _{PT}=1$.
The interesting point about the construction of state $\ket{\Psi_{T}}_{\pm}$ is the fact that it remains form invariant as we apply the time evolution operator, $\exp^{-i \hat{H}_{T} t}$. After  straightforward algebra, the time  evolved state  can be expressed as
\begin{widetext}
\bea
\ket{\Psi_{T}^{} (t)}_{+} &=&\dfrac{1}{\sqrt{c}} \ket{\uparrow}_{z} \otimes e^{-i \hat{H}_{PT} ^{} t} \ket{\psi}_{PT}^{} + 
                           \ket{\downarrow}_{z} \otimes   \hat{\mathcal{Q}} \ e^{-i \hat{H}_{PT}^{} t} \ket{\psi}_{PT}^{} ~.
\label{E3}
\eea
\end{widetext}
Similar evolution follows for the state $\ket{\Psi_{T}(t)}_{-}$. Now, it is clear from the form of the above state that the measurement of $\sigma_{z}$ performed on the ancilla degree of freedom prepared in this state followed by a post-selection of the outcomes, $\ket{\uparrow}_{z}$, results in an ensemble of states that simulates the desired \pt symmetric non-Hermitian dynamics. Hence, to conclude, the unitary evolution of $\ket{\Psi_{T}(t)}_{\pm}$ followed by quantum measurement protocol on the ancilla spin results in the simulation of desired non-unitary dynamics. 
%
%
Let the right eigenstates of the \pt symmetric Hamiltonian, $\hat{H}_{PT}$,  be represented by $\ket{\psi}^{k}_{PT}$ with corresponding eigenvalues given by $\{\epsilon_{k}\}$ where, $k$  takes integer values and labels all the allowed eigenvalues of $\hat{H}_{PT}^{}$. Further, it should be noted that similarity transformation does not change the eigenvalues, hence,  $\{\epsilon_{k}\}$ also represents the eigenvalues of $\hat{h}$ such that $ \ket{\psi}_{PT}= \hat{\mathcal{S}} \ket{\psi}$ where $\hat{h}  \ket{\psi} = \epsilon_{k}  \ket{\psi}$. 
We substitute $\ket{\psi}^{k}_{PT}$ in place of $\ket{\psi}_{PT}^{}$ in Eq.~\ref{E3} to obtain
 \begin{widetext}
\bea
\ket{\Psi_{T}^{k}(t)}_{+}^{} &=& \dfrac{e^{-i \epsilon_{k}  t}}{\sqrt{c}} \left( \ket{\uparrow}_{z} \otimes  \ket{\psi}^{k}_{PT} ~+~  \ket{\downarrow}_{z} \otimes   \hat{\mathcal{Q}} \ {\ket{\psi}^{k}_{PT}} \right). 
\label{E3p}
\eea
\end{widetext}
Similar evolution follows for the state $\ket{\Psi^k_{T}(t)}_{-}$. 
The above equation suggests that $\{\ket{\Psi_{T}^{k}(t)}_{+}, \ket{\Psi_{T}^{k}(t)}_{-}\}$ forms a complete set of two-fold degenerate eigenstates of the total embedding Hamiltonian $\hat{H}_{T}^{}$ with eigenvalues given by $\{\epsilon_{k}\}$. It should further be noted that a specific linear combination of the above degenerate states can be written as a product state of the  ancilla spin and the rest of the \pt symmetric degrees of freedom given by
\begin{equation}
    \ket{\chi_{T}^{}}_{\pm}^{k} = \ket{\pm}_{y} \otimes  \dfrac{1}{\sqrt{c}} \left(\mathbb{I}\mp i \hat{\mathcal{Q}}\right) \ket{\psi}_{PT}^{k}~.
\label{E6}
\end{equation}
Hence, if we think of the ancilla spin as a qubit which is coupled to bath degrees of freedom where the details of the couplings are given by the embedding Hamiltonian (Eq.~\ref{E1}),
  then the eigenstates given by Eq.~\ref{E3p}  and Eq.~\ref{E6} may be thought of as the bright state (qubit-bath entangled state)~\cite{Anushya} and   dark state~\cite{Mohit} respectively.
\section{\label{embedding N} Embedding of ``N" free non-hermitian spin-$1/2$}
%
Our strategy for embedding  a many-body \pt symmetric non-Hermitian system is the following. 

\begin{enumerate}

\item[(i)]  We identify an exactly solvable many-body Hermitian Hamiltonian with well-defined local degrees of freedom, which we call the seed Hamiltonian denoted by $\hat{h}$. 

\item[(ii)] We then perform a similarity transformation generated by $\hat{\mathcal{S}}$  as given by Eq.~\ref{E0}. More importantly, the $\hat{\mathcal{S}}$ is chosen such that each local  {{Hermitian}} degree of freedom is replaced by its \pt symmetric counterpart hence keeping the local character of the  $\hat{h}$ intact as we generate its \pt symmetric counterpart using Eq.~\ref{E0}. 

\item[(iii)] Once we obtain the desired $\hat{H}_{PT}^{}$, we follow the minimal embedding scheme detailed  in Sec.~\ref{embedding} which adds an ancilla spin-$1/2$ degree of freedom to the existing space of states of $\hat{H}_{PT}^{}$  to obtain an Hermitian embedding Hamiltonian given by $\hat{H}_{T}$ (see Eq.~\ref{E1}). 

\item[(iv)] After obtaining $\hat{H}_{T}^{}$, we expand it in terms of the local degrees of freedom identified in the seed Hamiltonian, $\hat{h}$. This provides us with a clear idea about the complexity of interactions induced by the embedding scheme among the local degrees of freedom which is needed for simulating the non-Hermitian dynamics governed by $\hat{H}_{PT}^{}$.  Hence, it provides a pathway  enabling extension of  experimental attempt in the context single body \pt symmetric case~\cite{Peng} to  the many-body \pt symmetric dynamics. 

\end{enumerate}

It should be noted that if the \pt symmetric model is exactly solvable, then the embedding Hamiltonian is also exactly solvable even though it looks quite complicated in terms of the local degrees of freedom owing to the emergent ``all to all" $q$-body interactions. It turns out that the eigenfunctions of the embedding Hamiltonian can be expressed in terms of the eigenfunctions  of the \pt symmetric Hamiltonian as we will show   later. 

In the present work, we consider a seed Hamiltonian which corresponds to a collection of $N$ number of free spin-$1/2$ degrees of freedom which are exposed to a uniform magnetic field of unit amplitude in arbitrary units pointing along $\hat{x}$, $\hat{h}= \sum_{i=1}^{N} \hat{\sigma_{i}^{x}}$. Next, we identify the operator, $\hat{\mathcal{P}} = e^{\theta(\hat{n}.\sigma_1)} \otimes e^{\theta (\hat{n}.\sigma_2)}\otimes.... \otimes e^{\theta (\hat{n}.\sigma_N)}$ which is a Hermitian operator. Here, each  $e^{\theta(\hat{n}.\sigma_{i})}$ acts on the Hilbert space of the corresponding $i^{th}$ spin considered in the seed Hamiltonian, where $\sigma_{i}$ is the  Pauli vector and $\hat{n}=( \sin \theta_{1} \cos \phi_{1},  \sin \theta_{1} \sin \phi_{1},  \cos \theta_{1})$ is  unit vector in three dimensions. It is straightforward to evaluate   the sum of inverse of eigenvalues of the operator $\hat{\mathcal{P}}$ given by $c = 2^{N} \cosh^{N} 2 \theta$. To obtain an interpretation of $\hat{H}_{T}$ in terms of the spin degrees of freedom, we expand $\mathcal{Q}$  in an infinite series as $\hat{\mathcal{Q}} = \sum_{m=0}^{\infty} (-1)^{m} \binom{1/2}{m} c^{-m+1/2} \  \hat{\mathcal{P}}^{2m-1} $. Expansion of $\mathcal{Q}$ allows us to write the explicit forms of operators $\hat{\mathcal{A}}$ and $\hat{\mathcal{B}}$ as
\bea
\hat{\mathcal{A}}&=&\sum_{i=1}^{N} \left[ (1/c) \ \hat{\mathcal{P}}  \hat{\sigma}_{i}^{x} \hat{\mathcal{P}}  +   \sum_{l,m=0}^{\infty} C_{lm}   \hat{\mathcal{P}}^{2 l} \ \hat{\sigma}_{i}^{x} \ \hat{\mathcal{P}}^{2 m}\right]\nn\\
\hat{\mathcal{B}} &=&  i \ \sum_{i=1}^{N}  \sum_{l=0}^{\infty} D_{l}  \left[ \hat{\mathcal{P}} ~ \hat{\sigma}_{i}^{x} ~ \hat{\mathcal{P}}^{2 l}  -  \hat{\mathcal{P}}^{2 l}\ \hat{\sigma}_{i}^{x} ~ \hat{\mathcal{P}}\right] ~,
\label{AB}
\eea
where $\hat{\sigma}_{i}^{x}=  \mathbb{I} \otimes \mathbb{I} \otimes \ldots \otimes \sigma_{i}^{x} \otimes \ldots \otimes \mathbb{I}$, $C_{lm}=\binom{1/2}{l} \binom{1/2}{m} (-c)^{-(l+m)}$ and $D_{l} = (-1)^{l} (c)^{-(l+1/2)} \binom{1/2}{l}$. Hence, upon substitution the total Hamiltonian, $\hat{H}_{T}$ in Eq.~(\ref{E1}) can be written in terms of a series  

\begin{widetext}
\bea
\hat{H}_{T} &=&   \mathbb{I}_{2\times 2} \otimes \left\{   \sum_{i=1}^{N} \ A_{0} \hat{\sigma}_{i}^{n} + \sum_{i=1}^{N} \ A_{1} \hat{\sigma}_{i}^{x} 
                         + \sum_{j\neq i}^{N} A_{2}  \hat{\sigma}_{i}^{x} \hat{\sigma}_{j}^{n} + \sum_{j\neq i \neq k}^{N} A_{3}  \hat{\sigma}_{i}^{x} \ \hat{\sigma}_{j}^{n} \hat{\sigma}_{k}^{n}+ \ldots \right\} \nn\\
                      &&    + ~ \sigma_{y} \otimes \left\{  \sum_{i=1}^{N} \ B_{1} \hat{\sigma}_{i}^{\perp x} + \sum_{j\neq i}^{N} B_{2} \hat{\sigma}_{i}^{\perp x} \hat{\sigma}_{j}^{n} +  
                           \sum_{j\neq i \neq k}^{N} B_{3}  \hat{\sigma}_{i}^{\perp x}  \hat{\sigma}_{j}^{n} \hat{\sigma}_{k}^{n}+ ....  \right\} ~,                           
\label{E7}
\eea
\end{widetext}
where, the constants  $A_{j}$ and $B_{j}$ can be determined by   substituting  $e^{ \theta(\hat{n}.\sigma_i)} = \cosh \theta~  \mathbb{I} +  \sinh \theta ~(\hat{n}.\sigma_i)$ in expression of $\hat{\mathcal{A}}$ and $\hat{\mathcal{B}}$ given by Eq.~\ref{AB}. Note that $\hat{\sigma}_{i}^{n}$ is the component of the spin operator along $\hat{n}$  for the $i^{th}$ non-Hermitian spin and $\hat{\sigma}_{i}^{\perp x}$ is component of spin operator perpendicular to $x$. We note that it results in  ``all to all"  $q$-body Hamiltonian ($q=1,...,N$) for the spin originating from the seed Hamiltonian. It also generates ``all to all"  $q$-body ($q=1,...,N+1$,  with $N \to$ spins of seed Hamiltonian and $1 \to$ ancilla spin) interaction terms between the ancilla spin and the spins of the seed Hamiltonian where all such interactions are anisotropic and  mediated only via $\sigma_{y}$ component of the ancilla spin. Thus, in order  to simulate even  a collection of $N$ free spins which constitutes one of simplest possible many-body \pt symmetric systems, one needs to engineer a very complex interaction network. In the limit, $\hat{n} \to \hat{x}$,  $\hat{H}_{T} \to \hat{h}$  owing to the fact that  $[\hat{h},\hat{\mathcal{Q}}]=0$ and hence $\hat{\mathcal{P}}(\hat{h}+\hat{\mathcal{Q}}\  \hat{h}\hat{\mathcal{Q}})\hat{\mathcal{P}}=\hat{h}$.
 %
\section{\label{embedding N=2} Explicit form of Embedding Hamiltonian for $N=2$  } 

For explicit demonstration of resultant $q$-body interaction generation, we take the simplest example of $N=2$ non-Hermitian spin halves. The seed Hamiltonian $\hat{h}$ is $\hat{h} = ( \mathbb{I}  \otimes \sigma_{x} + \sigma_{x} \otimes \mathbb{I})$ and we choose $ \hat{\mathcal{P}} = e^{\theta \sigma_{z}} \otimes e^{\theta \sigma_{z}} $ for the sake of simplicity. We can then write the total Hamiltonian $\hat{H}_{T}$ including the interacting terms 
\bea
   \hat{H}_{T} &=&  \mathbb{I}_{2\times 2} \otimes \left\{  
   \sum_{i=1}^{2} A_{1} \ \hat{\sigma}^{x}_{i} + \sum_{j\neq i}^{2} A_{2} \  \hat{\sigma}_{i}^{x} \hat{\sigma}_{j}^{z} 
   \right\}  
                         \nn\\
                         && +~ \sigma_{y} \otimes \left\{  \sum_{i=1}^{2} B_{1} \ \hat{\sigma}_{i}^{y} + \sum_{j\neq i}^{2} B_{2} \ \hat{\sigma}_{i}^{y} \hat{\sigma}_{j}^{z} \right\}
\label{E8}
\eea
%
%
%
%
\begin{figure}[htb!]
\centering
\includegraphics[width=8cm, height=6cm]{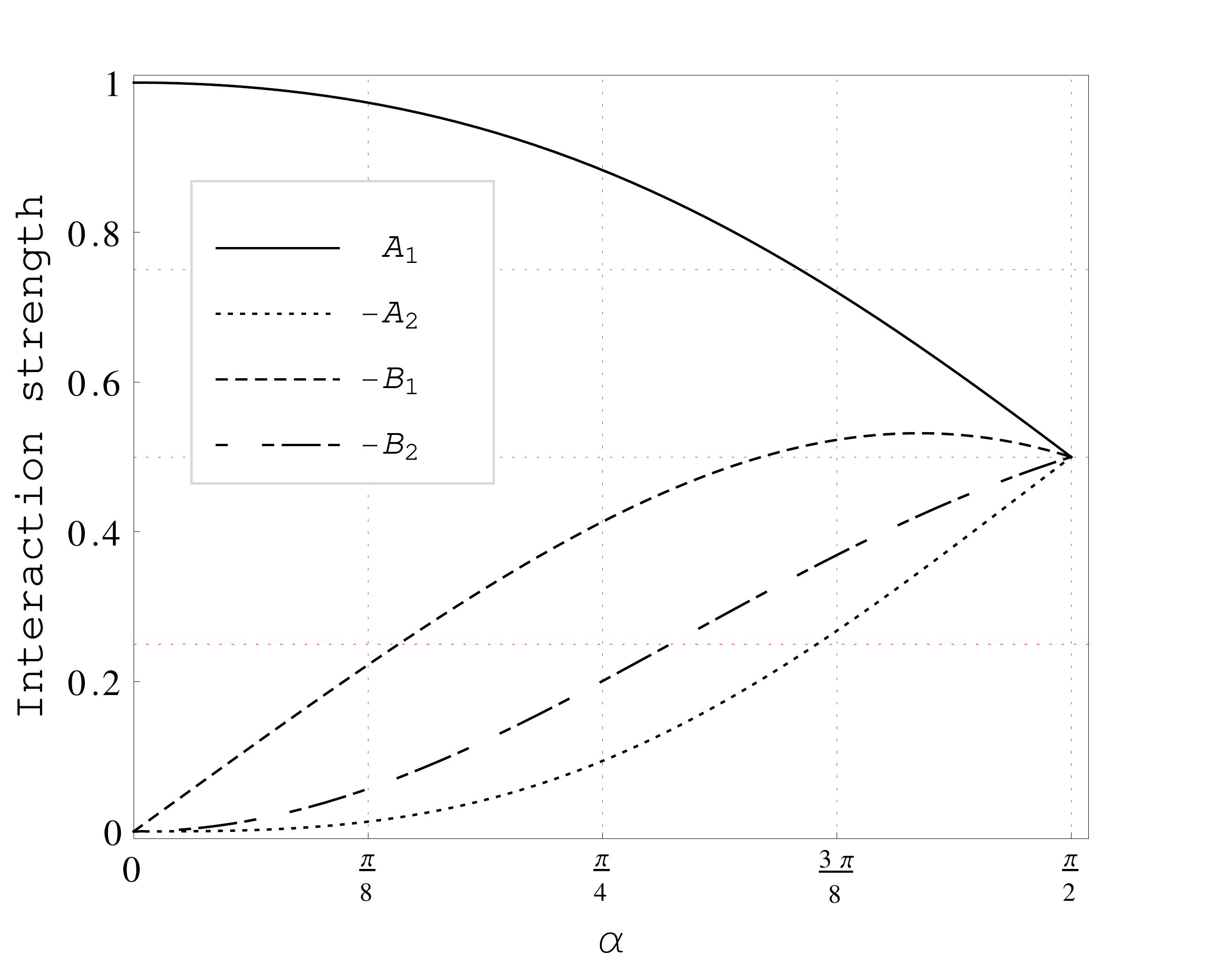}
\caption{Interaction strength parameters ($A_{1}, -A_{2}$, $-B_{1}$ and $-B_{2}$)  plotted as a function of parameter $\alpha$ defined using $\theta = (1/2) \tanh^{-1} (\sin \alpha)$. 
}
\label{f2}
\end{figure}
Note that the other interaction coefficients in Eq.~\ref{E7} vanish as the choice of operator, $\hat{\mathcal{P}}$ decides the non-zero coefficients. The non-zero coefficients  are plotted in the Fig.~\ref{f2}.
The term containing $A_{1}$ represents one-body term in the total Hamiltonian $\hat{H}_{T}$  and decreases as we tune $\alpha$ from $ 0 \to\pi/2$.
Two-body terms (with coefficients $B_{1}$ and $A_{2}$) of $\hat{H}_{T}$ grow as we tune $\alpha$ from $ 0 \to\pi/2$. The three-body term (involving $B_2$), which was not there in the Hamiltonian to begin with (at $\alpha=0$) also grows as we tune $\alpha$ from $ 0 \to\pi/2$.  As we have chosen $\hat{n}= \hat{z}$, all these  coefficients depend only on $\alpha$.  It is interesting to note that as $\alpha \to \pi/2$, the relative magnitude  of the one-body term ($A_1$) matches the amplitude corresponding to all the other interaction terms ($A_2$, $B_1$, $B_2$) which essentially means that interactions can not be treated perturbatively and one is in strong correlation regime. We will discuss later that even for large $N$,  the limit, $\alpha \to \pi/2$  remains relevant from the point of view of non-trivial correlations between spins. 

%
\section{\label{entanglement} Embedding induced entanglement in eigenstates} 
%

From Eq.~\ref{E7}, it is evident that the procedure of embedding generates complicated interactions even though our seed Hamiltonian has free spins. This  leads to entanglement generation  among the spins when prepared in the eigenstate of  embedding Hamiltonian $\hat{H}_{T}$. For exploring this, we  work in the basis given by Eq.~\ref{E6} where the eigenstates of $\hat{H}_{T}$ are given as a product state of the ancilla spin and rest of the spins which could be thought of as bath-spins. 
  This is a convenient basis to work with as we are interested in studying entanglement between the bath spins only and hence we want to avoid any contribution to entanglement induced by the ancilla spin. In particular, we focus on the ground state of $\hat{H}_{T}$ which can be obtained from the ground state of the seed Hamiltonian $\hat{h}= \sum_{i=1}^{N} \hat{\sigma_{i}^{x}}$ given by $\ket{\psi}^{0}= \bigotimes^{N} \ket{\downarrow}_{x}$ and corresponds to all spins pointing along $-x$ direction. The ground state for the corresponding \pt symmetric Hamiltonian can be obtained via the similarity transformation given by $\ket{\psi}_{PT}^{0} = \hat{\mathcal{P}} \ket{\psi}^{0} = e^{\theta(\hat{n}\cdot \sigma_1)} \ket{\downarrow}_{x} \otimes e^{\theta (\hat{n} \cdot\sigma_2)}\ket{\downarrow}_{x} \otimes \ldots \otimes e^{\theta (\hat{n} \cdot \sigma_N)}\ket{\downarrow}_{x}$. And, hence  the corresponding doubly degenerate ground states of $\hat{H}_{T}$ can be written as  $\ket{\chi_{T}}_{\pm}^{0} = \ket{\pm}_{y} \otimes (1/\sqrt{c}) (\hat{\mathcal{P}} \mp  i  \hat{\mathcal{P}}  \hat{\mathcal{Q}}) \ket{\psi}^{0}$.   
  We focus on the states $(1/\sqrt{c}) (\hat{\mathcal{P}} \mp  i  \hat{\mathcal{P}} ~ \hat{\mathcal{Q}}) \ket{\psi}^{0}$ which solely represent the bath spins and carry the information about  entanglement between bath spins. We note that the operators  $\hat{U}_{\pm} =(1/\sqrt{c}) (\hat{\mathcal{P}} \pm i  \hat{\mathcal{P}} \hat{\mathcal{Q}})$ are unitary and their action on the bath spins is not local and hence it is responsible for generation of entanglement. 
  
  Furthermore, the operators $\hat{\mathcal{P}}/\sqrt{c}$ and $\hat{\mathcal{P}} \hat{\mathcal{Q}}/\sqrt{c}$ commute with each other and we can define a unitary transformation $U_{\hat{\mathcal{P}}}$ which diagonalizes both the operators simultaneously, i.e.,   $(1/\sqrt{c}) (\hat{\mathcal{P}}\pm i  \hat{\mathcal{P}}  \hat{\mathcal{Q}})=U^{\dagger}_{\hat{\mathcal{P}}}(\hat{\mathcal{D}}_{P} \pm i  \hat{\mathcal{D}}_{P Q }) U_{\hat{\mathcal{P}}}$. It is important to note that operator $U_{\hat{\mathcal{P}}}$  is a tensor product of local (each acting on individual spins) unitary operators as each operator $e^{\theta (\hat{n}\cdot\sigma)}$ in the definition of operator $\hat{\mathcal{P}}$ can be written as $e^{\theta (\hat{n}\cdot \sigma)}= u^{\dagger} e^{\theta \sigma_{z}} u$, implying $U_{\hat{\mathcal{P}}} = \bigotimes^{N} u$. Furthermore,  $\hat{\mathcal{D}}_{P} \pm i  \hat{\mathcal{D}}_{P Q }$ are diagonal unitary operators which are responsible for generation of entanglement as the rest of it, i.e. $U_{\hat{\mathcal{P}}}$,  is a local operator. Each element of diagonal unitary operators $\hat{\mathcal{D}}_{P} \pm i  \hat{\mathcal{D}}_{P Q }$  should be of the form $e^{i \gamma}$ due to uni-modular property of diagonal unitary operator and hence the eigenvalues of operator $\hat{\mathcal{D}}_{P}$ can be written in decreasing value sequence as: $ e^{N \theta}/\sqrt{c},e^{(N-2) \theta}/\sqrt{c},e^{(N-4) \theta}/\sqrt{c}, \ldots,e^{-N \theta}/\sqrt{c}$.  We next discuss some of the limiting cases in the parameter space. 
\begin{figure}[htb!]
\centering
\includegraphics[width=7.5cm, height=6cm]{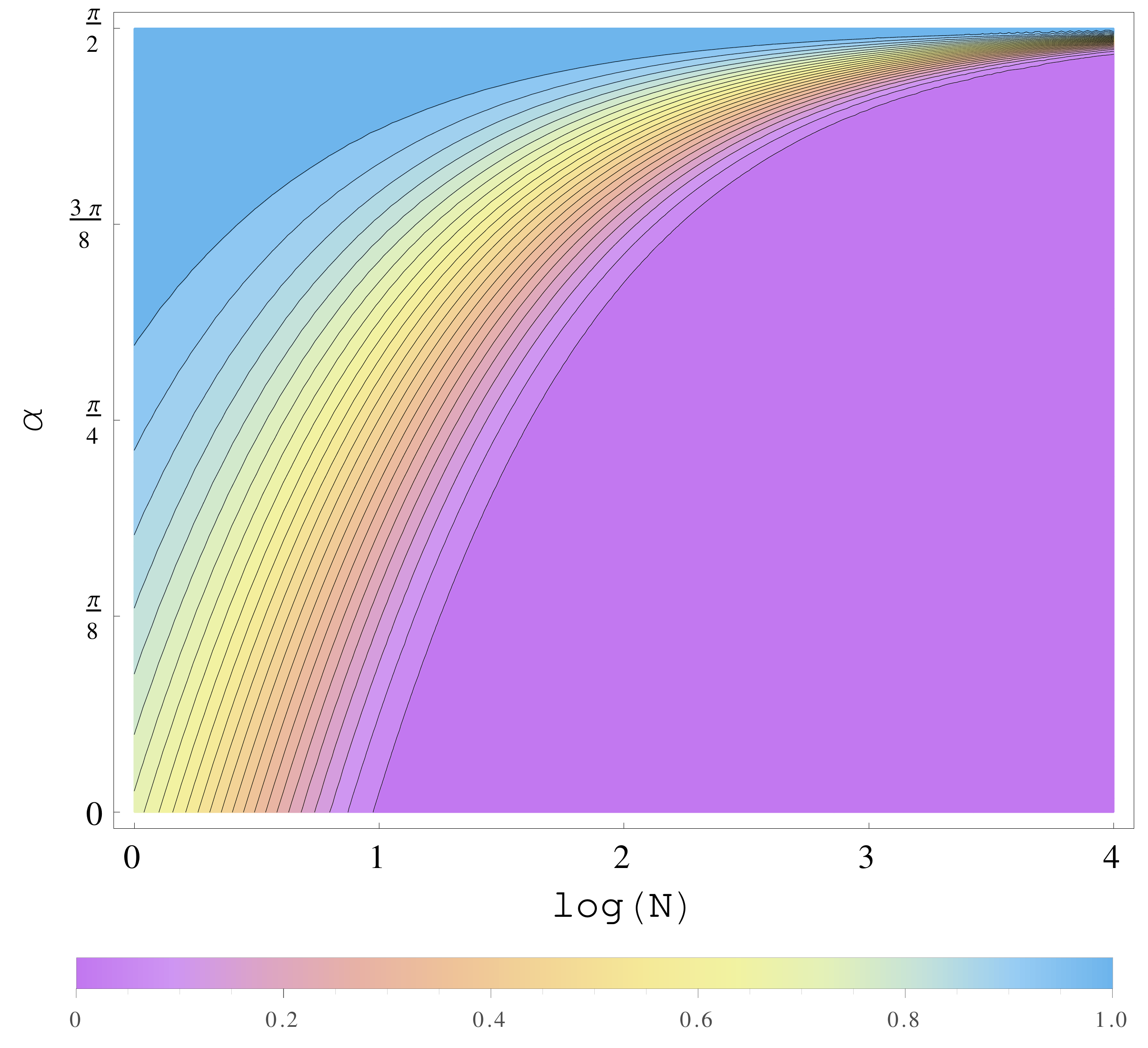}
\caption{The largest eigenvalue, $e^{N \theta}/\sqrt{c}$ of the operator, $\hat{\mathcal{D}}_{P}$ plotted  in the plane of $\alpha-\log(N)$. 
}
\label{f3}
\end{figure}
\begin{itemize}
\item 
{\underline{$\alpha \rightarrow \pi/2$  (very large $\theta$)  limit :}}

In this limit, only the first eigenvalue of $\hat{\mathcal{D}}_{P}$ given by $e^{N \theta}/\sqrt{c}$ is dominant and given by 
\begin{equation}
\frac{e^{N \theta}}{\sqrt{c}} = \frac{1}{(1+ e^{-4 \theta})^{N/2}} = \cos \gamma \rightarrow 1,
\label{En1}
\end{equation}
where we have used the explicit expression for  $c$. Similarly, corresponding eigenvalue of operator $\hat{\mathcal{D}}_{P Q }$ given by  $e^{N \theta} \sqrt{-1 + c ~ e^{-2 N \theta}}/\sqrt{c}= \sin \gamma \rightarrow 0$ in this limit. Rest of the eigenvalues of $\ \hat{\mathcal{D}}_{P} \to 0$ as we approach $\alpha \rightarrow \pi/2$, owing to an exponential decay.  This allows us to write the states, $ \ket{B}_{\pm}^{0} = (1/\sqrt{c}) (\hat{\mathcal{P}} \pm  i  \hat{\mathcal{P}}  \hat{\mathcal{Q}}) \ket{\psi}^{0}$ in $\alpha \rightarrow \pi/2$ limit as
\begin{equation}
\ket{B}_{\pm}^{0} \approx  U^{\dagger}_{\hat{\mathcal{P}}} \left[\hat P_{\uparrow} \pm  i (\mathbb{I}- \hat P_{\uparrow} )\right]  \ket{\bar\psi}^{0} 
\label{En2}
\end{equation}
where $\hat P_{\uparrow}= \otimes^{N} \ket{\uparrow}_{z} \bra{\uparrow}_{z}$ is a direct product of local projection operators and $ \ket{\bar \psi}^{0}  = U_{\hat{\mathcal{P}}}   \ket{\psi}^{0}$. The above expression shows that $\ket{B}_{\pm}^{0}$ is a linear combination of two many-body state each of which can be expressed as direct product of local spin degrees of freedom which ensures finite entanglement between the bath spins. The degree of entanglement in this state depends on the interplay of $N$ and $\alpha$. As $\hat P_{\uparrow}$ and $ \ket{\bar\psi}^{0} $ are both local in spins, hence the action of $\hat P_{\uparrow}$ on this state results in a state whose amplitude scales as the overlap of states for one of the spin in $ \ket{\bar\psi}^{0} $ with its corresponding $\ket{\downarrow}_{x}$ state raised to the power of $N$. Hence, for a given $N$ one can optimize $\hat{n}$ such that this overlap raised to the power of $N$ remains finite. This fact plays an crucial role in large $N$ limit while for a a small value of $N$ it may not of that much consequence as far in determining the entanglement content of the state is concerned. And, this ensures that the state $\ket{B}_{\pm}^{0}$ indeed stays as a linear combination of two many-spin states  each having a non-vanishing amplitude hence leading to finite entanglement.

 \item 
 {\underline{Intermediate  $\alpha$ (finite $\theta$) and large $N$ limit :}}
 
\begin{figure}[b!]
\centering
\includegraphics[width=3.5cm, height=3.5cm]{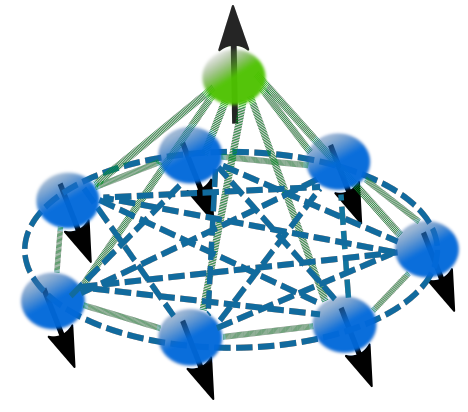}
\caption{ A schematic   of the central spin model where the interaction with the central spin is shown in solid green lines and interaction between bath spins is shown in dashed blue lines. }
\label{f0}
\end{figure}
 In this case, even the first eigenvalue of Eq.~\ref{En1} vanishes and the operator $\hat{\mathcal{D}}_{P}$ becomes a null operator while $\hat{\mathcal{D}}_{P Q }$ reduces to an Identity operator. Hence, the state $\ket{B}_{\pm}^{0}$ becomes  separable state in this limit. 

\end{itemize}
For any other choice of the parameters, there is an interplay between $N$ and $\alpha$ which acts as a deciding factor governing the degree of entanglement. This  is  demonstrated in  Fig.~\ref{f3}, where the largest eigenvalue  $e^{N \theta}/\sqrt{c}$ given in Eq.~\ref{En1} is plotted in the plane of $\alpha-N$. From the above analysis, we can conclude that presence of entanglement is primarily controlled by interaction parameter $\alpha$ and the parameter $\hat{n}$ only helps in optimizing it for a given $N$.   


\section{\label{central-spin} Central Spin Model and orthogonality catastrophe} 
Following the line of thought presented below Eq.~\ref{E6}, one can think of the embedding  Hamiltonian in Eq.~\ref{E7}   analogous to a ``central spin model" with the extra spin half degree of freedom (added for the purpose of embedding)  and referred to as the ancilla depicting the ``central spin". The interesting point is that the  embedding Hamiltonian contains ``all to all" interaction terms however, all spins (other than the ancilla) remain indistinguishable from the point of view of the spin-spin interaction and this is what makes the central spin distinct from the rest. A schematic representation of emergent central spin model is shown in Fig.~\ref{f0}.

As the eigenstates, $\ket{\chi_{T}^{}}_{\pm}^{k}$ are product states of the ancilla spin and the spin-bath, these states are analogous to the dark states observed in light-atom interaction. This has been discussed recently in the context of central spin model~\cite{Anushya}. Now, we will show that the ancilla state in our model, when prepared in $\ket{\chi_{T}^{}}_{\pm}^{k}$   serves as a robust quantum information storage unit owing to protection against spin-flipping perturbation mediated by the interaction acting on the ancilla (or central qubit). We first  introduce an additional term in the Hamiltonian ($\hat{H}_{m}$) pertaining to application of a magnetic field of strength $m_y$ pointing along $\hat{y}$ acting only on the ancilla (central spin) whose effect on the spectrum of $\hat{H}_{T}$ is to lift the degeneracy of the many-body eigenstates $\ket{\chi_{T}^{}}_{\pm}^{k}$ and split them by $2 m_{y}$ units of energy. Of course, $\ket{\chi_{T}^{}}_{\pm}^{k}$ continues to be eigenstate of the total Hamiltonian given by $ \hat {H}_{total}=\hat{H}_{T}^{}+\hat{H}_m^{}$ since $ \left[\hat{H}_{T}^{}, \hat{H}_m^{}\right]=0$. Then, the probability of flipping the ancilla (central spin) due to a transverse field $m_{z} \hat{z}$ acting on it, in the small field limit will be proportional to     $|^{~~k}_{~~-}  \bra{\chi_{T}^{}}_{} (\sigma_{z} \otimes \mathbb{I})  \ket{\chi_{T}^{}}_{+}^{k}|^{2}$, which inturn is proportional to $|^{~~k}_{~~+}\braket{B_{}^{}| B}_{-}^{k}{} |^{2}$. Hence,  orthogonality of $\ket{B}_{\pm}^{k}$ could result in large suppression of such spin-flip processes which could be attributed to  Anderson's orthogonality catastrophe~\cite{Anderson}. We next show that the interplay of the interaction parameter $\theta$ ($\theta=0$ being the free limit) and the size of the bath $N$ together can conspire to give orthogonality of $\ket{B}_{-}^{k}$ and $\ket{B}_{+}^{k}$ for arbitrary values $N$ but this works only for the ground state manifold (i.e., the $k=0$ space).  

The orthogonality implies that for $\hat{H}_m=0$, the two-fold degenerate $k=0$ eigenspace of $\hat{H}_{T}$, hosts state which are maximally entangled between the  ancilla  (central spin)
 and the bath given by $\ket{\Psi_{T}^{k=0}}_{\pm}$ in Eq.~\ref{E3p} such that the von-Neumann entropy of the reduced density operator for the central spin equals $\ln 2$. In general the state given in Eq.~\ref{E3p} for all $k's$ represents states that are analogous to the bright state~\cite{Anushya} observed in light-atom interaction. Hence we conclude that  the eigenstates of $\hat{H}_{T}$ in absence of $\hat{H}_m$ could be either dark or bright but a maximally entangled bright state exists only in the ground state manifold. Now will analytically explore this orthogonality for the $k=0$ subspace.

To demonstrate  orthogonality of  $\ket{B}_{+}^{0}$ and $\ket{B}_{-}^{0}$ we evaluate the following
\bea
 ^{~~0}_{~~-} \scx{B}{B}_+^0 &=& -1 +  \dfrac{2}{c} \Big[ {}^{~~0}\me{\hat{\mathcal P^2}}{\psi}{\psi}^0
  \nonumber \\ 
  && 
  +~ i ^{~~0} \me{{\hat{\mathcal{P}}}{\hat{\mathcal{Q}}}{\hat{\mathcal{P}}}}{\psi}{\psi}^0
   \Big]~.
    \label{Sp1}
\eea
We know that $\hat{\mathcal{P}}/\sqrt{c} = U^{\dagger}_{\hat{\mathcal{P}}}  {\hat{\mathcal{D}}}_{\mathcal{P}}^{}  U^{}_{\hat{\mathcal{P}}}  $ and 
 $\hat{\mathcal{P}} \hat{\mathcal{Q}}/\sqrt{c} = U^{\dagger}_{\hat{\mathcal{P}}} ~ \hat{\mathcal{D}}_{P Q} ~U_{\hat{\mathcal{P }}} $. Thus, Eq.~\ref{Sp1} can be rewritten as 
\bea
 ^{~~0}_{~~-} \scx{B}{B}_+^0 &=& -1 +  {2} \Big[ {}^{~~0}\me{\hat{\mathcal D_{P}^2}}{\bar\psi}{\bar\psi}^0
  \nonumber \\ 
  && 
  +~ i ^{~~0} \me{{\hat{\mathcal D_{P}}} {\hat{\mathcal D_{PQ}}} }{\bar\psi}{\bar\psi}^0
   \Big]~.
    \label{Sp2}
\eea
%
Let us consider the following limiting cases :
\begin{itemize}
\item 
{\underline{Intermediate $\alpha$  and large $N$ limit :
}}

 In this limit, the operator $\hat{\mathcal{D}}_{P}$ reduces to a null operator and hence the overlap in Eq.~\ref{Sp2}  reduces to $^{~~0}_{~~-} \scx{B}{B}_+^0 = -1$. This 
  implies that the two states given by $\ket{B}_{\pm}^{0}$ and not different and are infact the same up to a phase of $\pi$.  In fact, for a given $\alpha$,
    we can always increase the number of spins in the bath proportionately in order to make the states $\ket{B}_{\pm}^{0}$ co-linear with respect to each other. 
\item 
{\underline{$\alpha \rightarrow \pi/2$ (large $\theta$) and finite $N$ limit :}}
 
In this case,   the operators $\hat{\mathcal{D}}_{P}$ reduce to projection operator,  $\hat P_{\uparrow}= \otimes^{N} \ket{\uparrow}_{z} \bra{\uparrow}_{z} $ which implies   $\hat{\mathcal{D}}_{P}^{2} = \hat{\mathcal{D}}_{P}$ and $\hat{\mathcal{D}}_{P Q} = \mathbb{I}- \hat P_{\uparrow}$. The product $\hat{\mathcal{D}}_{P} \hat{\mathcal{D}}_{P Q}$ represents  null operator. The  non-trivial (second) term  on right hand side of Eq.~\ref{Sp2} can be expressed as
$|_z\me{u}{\uparrow}{\downarrow}_x |^{2N}$
 where $u$ is unitary operator such that $e^{\theta (\hat{n} \cdot \sigma)}= u^{\dagger} e^{\theta \sigma_{z}} u$. Recall that, the operator $u$ is a function of $\theta_{1}$ and $\phi_{1}$ which defines the unit vector $\hat{n}$. In terms of $\theta_{1}$ and $\phi_{1}$, we can write it as 
\bea
|_z\me{u}{\uparrow}{\downarrow}_x |^{2N} &=&
  \dfrac{(1- \cos \phi_{1} \sin \theta_{1})^{N}}{2^{N}}
    \label{Sp3}
\eea
\begin{figure}[htb!]
\centering
\includegraphics[width=8cm, height=7cm]{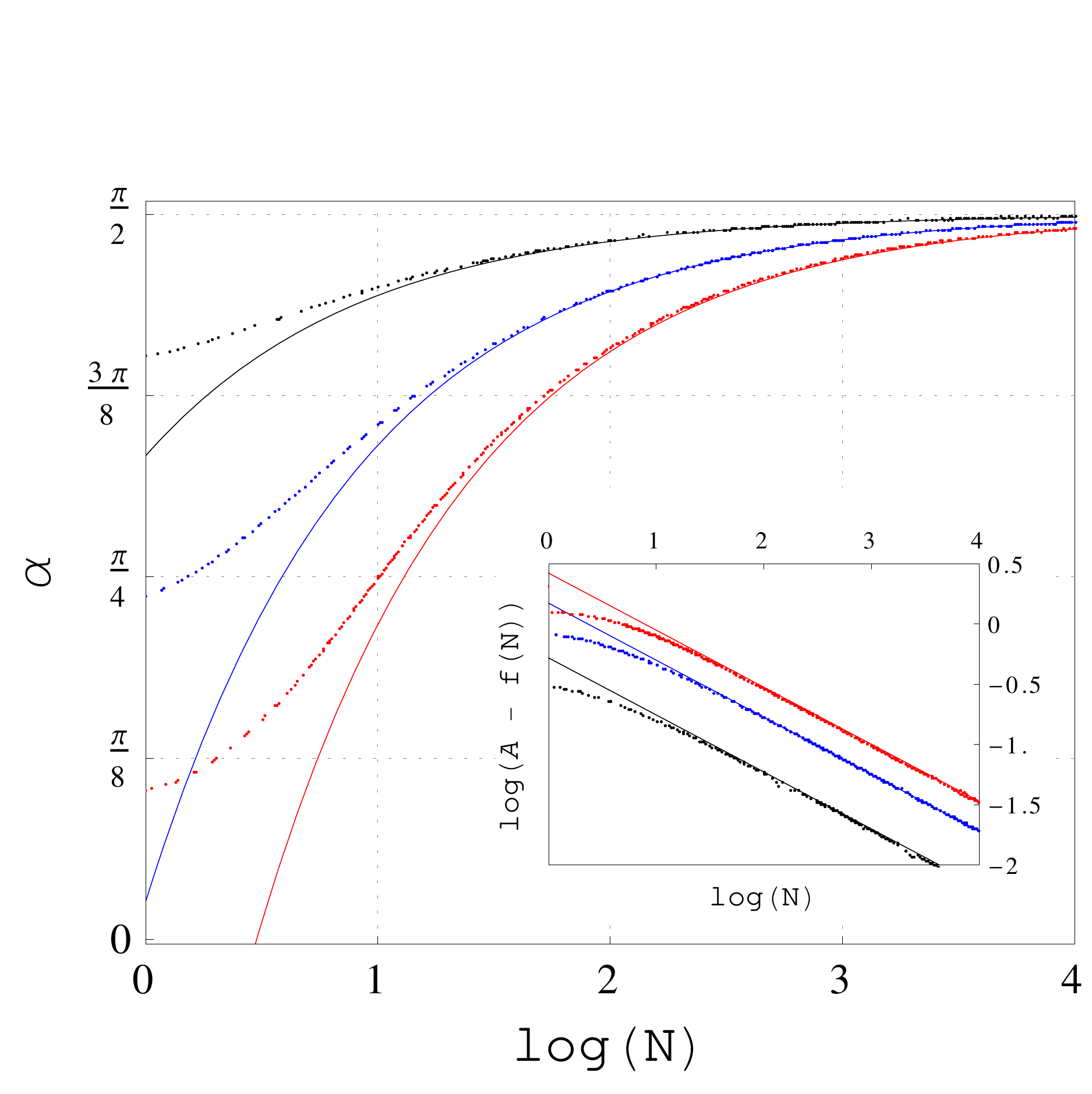}
\caption{Contours of fixed values of overlap of two states  $|^{~~0}_{~~-} \scx{B}{B}_+^0|^2$ is fitted with power law functions in the $\alpha- \log(N)$ plane. The data is fitted with the power law  function of the  form $f(N) = A - B N^{- \gamma}$ with $\gamma=0.473$. Data points are shown as dots of same colour for contours of fixed $|^{~~0}_{~~-} \scx{B}{B}_+^0|^2$  corresponding to 
  0.09 (black) ,   $0.54$
(blue) and   $0.90$ (red).    
The parameter $B$ has values of $2.647$, $1.491$ and $0.524$ for overlap 0.90, 0.54 and 0.09 respectively. The parameter $\hat n$ for all the three curves has been parameterized by  defining $\theta_1 = \pi/2$ and $\phi_1 = 2 \sin^{-1} (\sqrt{2^{-1/N}})$ and $A_{}$ takes value $1.572$.  
We plot $\log(A_{}-f(N))$ as a function of $\log(N)$ in the inset clearly depicting the emergence of a power law dependence at large $N$.
}
\label{f5}
\end{figure}
To verify if the above relation is satisfied for some values of $\theta_1$ and $\phi_1$, let us choose $\theta_{1} = \pi/2$, i.e. $\hat{n}$ lies in the $x$-$y$ plane, which implies  $ |_z\me{u}{\uparrow}{\downarrow}_x |^{2N}  = (\sin {\phi_{1}/2})^{2 N}$. For orthogonality of $\ket{B}_{\pm}^{0}$, we require $ (\sin \phi_{1}/2)^{2 N} = 1/2$. This leads to $N$ dependence of parameter given by $\phi_{1} = 2 \sin^{-1}( \sqrt{2^{-1/N}})$. It is worth noting that the parameter regime of orthogonality of $\ket{B}_{\pm}^{0}$  has overlap with parameter regime where the bath states depict finite entanglement (see Sec.~\ref{entanglement}). Hence, the $\alpha \rightarrow \pi/2$ for finite $N$ is an interesting limit as the orthogonality of these two many-body state seems to be directly related to the presence of finite entanglement between the bath spins which in turn is indicative of an orthogonality catastrophe which we will discuss shortly.  
\end{itemize}

From Fig.~\ref{f5}. it is interesting to note that the overlap of states depicting the bath,   $|^{~~0}_{~~-} \scx{B}{B}_+^0 |^2$ follow   universal scaling  in $\alpha-\log(N)$ plane for the values corresponding to an optimized choice of $\theta_{1}$ and $\phi_{1}$ which ensure orthogonality of the states $\ket {B}_{-}^{0}$ and $\ket{B}_{+}^{0}$ in the limit, $\alpha \rightarrow \pi/2$  discussed above. The inset of  Fig.~\ref{f5} clearly demonstrates the emergence of a universal power law scaling in $N$ for  the large $N$ limit, i.e., the power law exponent obtained from all the contours of fixed  $|^{~~0}_{~~-} \scx{B}{B}_+^0 |^2$ have the same value.

Henceforth, we shall use the term orthogonality catastrophe  to imply the orthogonality of $\ket{B}_{-}^{0}$, $\ket{B}_{+}^{0}$ when the states are orthogonal due to reasons different from the  possibilities given below~\cite{chalker}

\begin{itemize}

\item 
 The likelihood that two randomly  chosen  unit  vectors (let's say, $\ket{\Psi}$ and $\ket{\Phi}$)  from  an $2^{N}$-dimensional  complex  vector  space $\mathbb{C}^{2^{N}}$  will exhibit  orthogonality as we increase the dimension of the Hilbert space scales as $f_{1}=\overline{|\braket{\Psi|\Phi}|^{2}}= 1/2^{N}$.

\item
The overlap between two product states   of $N$ spin halves, $\ket{S_{1}} = \otimes^{N} \ket{s_{1}}$ and $\ket{S_{2}}= \otimes^{N} \ket{s_{2}}$ is given by  $f_{2}=|\braket{S_{1}|S_{2}}|^{2}= |\braket{s_{1}|s_{2}}|^{2 N} = (\cos \beta)^{2 N} $ assuming all spins are pointing in the same direction in a given product state is exponentially small in $N$,  $\ket{s_{1}}$ and $\ket{s_{2}}$. 
\end{itemize}

\begin{figure}[b!]
\centering
\includegraphics[width=7cm, height=6cm]{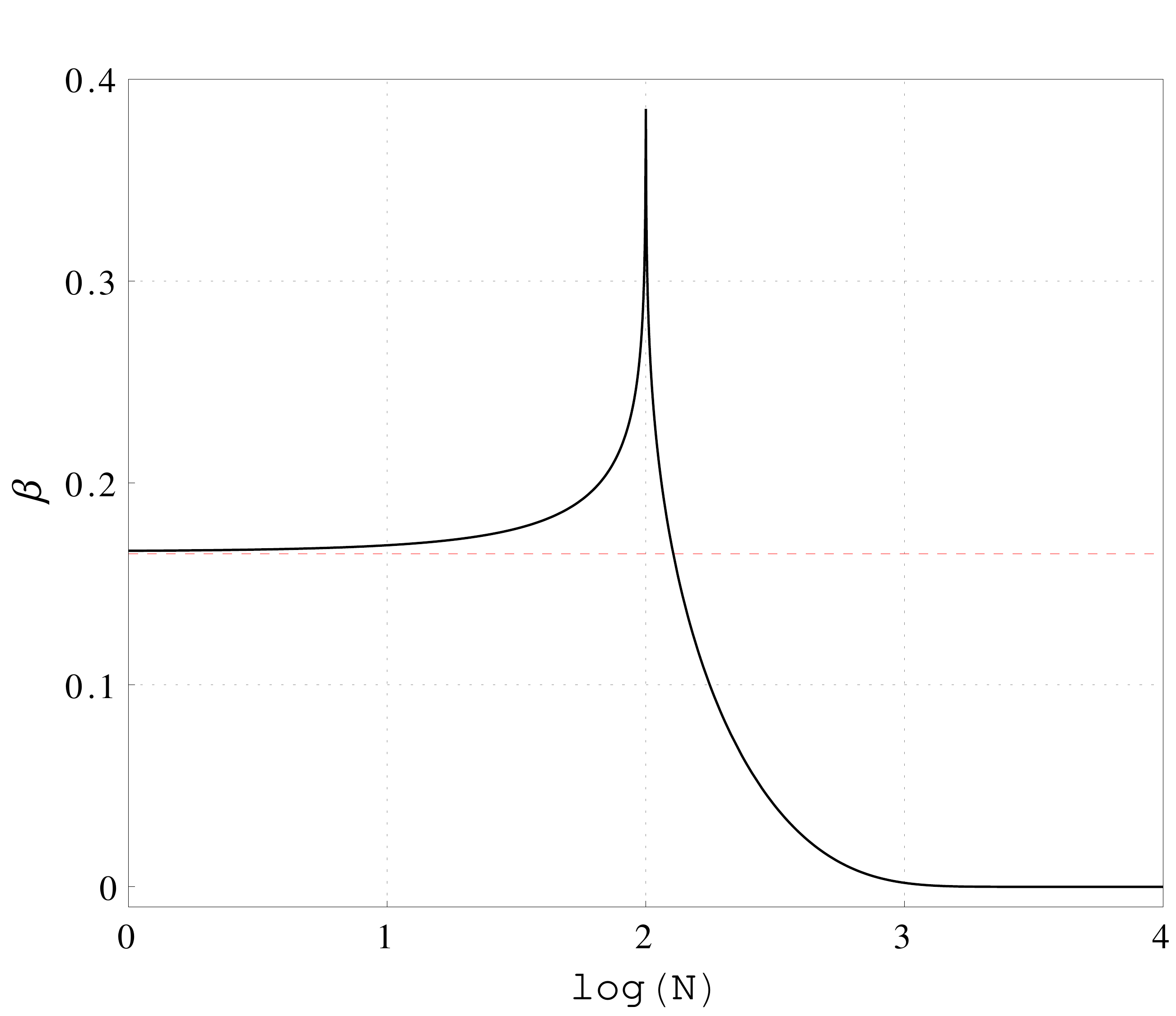} 
\caption{The parameter $\beta$ plotted as a function of  $\log(N)$. Dashed horizontal line is the guide to the eye to identify the domain of $N$ for which the functions $f_{2}$ reduces to $f_{3}$.  }
\label{f4}
\end{figure}

We further intend to demonstrate (numerically) under which conditions $f_3=|^{~~0}_{~~-} \scx{B}{B}_+^0 |^2$   could be distinct from $f_1$ and $f_2$. 
For this reason, we consider the limit, $\alpha \to \pi/2$   which we know (from the preceding discussion) gives rise to eigenstates that are distinct from a product state.   In this limit, $f_{3} $ takes the form $ f_3 = (-1 + 2 (\sin \phi_{1}/2)^{2 N})^{2}$. 
 For the sake of  numerical analysis,  we take $\theta_1= \pi/2$ and $\phi_{1} = 2 \sin^{-1}( \sqrt{2^{-1/N}}) | _{N=100}$ which ensures orthogonality of states  $\ket{B}_{\pm}^{0}$ for $N=100$. 
 For this choice of parameters,  $f_{3} = (-1 + 2^{1-(N/100)})^{2}$     which is clearly distinct from $f_1=1/2^{N}$ in the vicinity of $N=100$. Now, it remains to be checked if $f_3$ is indeed distinct from $f_2$ for the above choice of parameters. For $N \gg 100$, it is obvious that $f_{2}$ will have overlap with 
  $f_{3}$   as the bath state reduces to a product state in that limit. However, in neighbourhood of $N=100$, we can check if there is a finite overlap by equating $f_{2}$ to $f_{3}$ and solving  for the parameter $\beta$. Now, if we plot $\beta$ as a function of $N$, then we expect that  $\beta$ will stay constant over a domain of values of $N$ (i.e., it is independent of $N$) where the  functions $f_{2}$ and $f_{3}$ are equal.  A plot of $\beta$ obtained as a function of $N$ in given in Fig.~\ref{f4}.  It is evident that $f_{2}$ and $f_{3}$ are never equal in the vicinity of $N=100$, rather they deviate strongly from each other. As we go to larger values of $N$, say $N=100 \to 1000$,  $f_{3}$ reduces to $f_{2}$.
  This proves that the origin of the orthogonality of  $^{~~0}_{~~-} \scx{B}{B}_+^0 $ in the limit $\alpha\to \pi/2$ can    indeed be associated with the physics of orthogonality catastrophe i.e., it is driven by presence of finite entanglement.

   It is important to note that $N=100$ is chosen to demonstrate the existence of the orthogonality catastrophe which is 
arising from correlation in the bath spin but such an orthogonality can be organized for any large $N$  provided we choose the rest of the parameters ($\alpha$, $\phi_{1}$ and $\theta_{1}$) appropriately. 
\section{\label{discussion} Discussion and conclusion}

\pt symmetric systems and their physical realization has been largely carried out using classical optics platforms.  However, 
in recent times, there has been some progress  in realizing \pt symmetry in the quantum domain. For instance, 
in the experiment reported in Ref.~\onlinecite{Klauck}, a \pt symmetric two-photon system was realized while the experiment result reported in Ref.~\onlinecite{Li} was related to the observation of \pt symmetry breaking transitions in an optical dipole trap of ultra cold $^6 {Li}$ atoms. Hence, it is likely that complex \pt symmetric quantum many-body systems becomes 
 realizable in the near future.  
 
 On the  theoretical front, there are very few proposals which provide  a well-defined route towards realizing \pt symmetric quantum systems. The idea of embedding is one of the few ideas  which has been proposed~\cite{Ueda, Ray,Kumar}  and tested experimentally~\cite{Peng} for a one-photon and a two-photon system.  For the first time, we extend these ideas of  embedding to the domain of many-body systems. Our study provides  a close view of the complexity of embedding induced emergent interactions between the  local \pt symmetric degrees of freedom.

A spin off of our study is the ``central spin model'' which emerges from the reinterpretation of the embedding Hamiltonian where the extra degree of freedom added to the system for the purpose of embedding acts as the central spin. We show that the central spin can be prepared in a dark state which enjoys protection against spin-flipping perturbations (and hence from decoherence) owing to an orthogonality catastrophe. Hence, our embedding Hamiltonian poses an interesting example of  a viable model for a qubit (central spin) which enjoys protection from decoherence owing to its coupling to bath spins.


\section*{Acknowledgements}
AVV would like to thank the Council of Scientific and Industrial Research (CSIR), Govt. of India for financial support.  S.D. would like to acknowledge the MATRICS grant (MTR/ 2019/001 043) from the Science and Engineering Research Board (SERB) for funding. 


\end{document}